\documentclass[aps,pra,twocolumn,showpacs,nofootinbib]{revtex4-1}
\usepackage{amssymb}
\usepackage{color}
\usepackage{graphicx}
\usepackage{bbold}
\usepackage{bm}
\usepackage{amsmath}
\usepackage{amsfonts}
\usepackage{amssymb}
\usepackage{braket}
\usepackage{units}
\usepackage{csquotes}
\usepackage{upgreek}
\usepackage{xcolor}
\usepackage{fancyhdr}

\begin{document}

\title{Ultraviolet Fabry-Perot cavity with stable finesse under ultrahigh vacuum conditions}

\author{Jonas Schmitz, Hendrik M. Meyer, and Michael K\"ohl}

\affiliation{Physikalisches Institut, Universit\"at Bonn, Wegelerstrasse 8, 53115 Bonn, Germany}

\begin{abstract}
We have constructed an apparatus containing a linear ion trap and a high-finesse optical cavity in the ultraviolet spectral range. In our construction, we have avoided all organic materials inside the ultrahigh vacuum chamber. We show that, unlike previously reported, the optical cavity does not degrade in performance over a time scale of 9 months.
\end{abstract}

\maketitle

\section{Introduction}
Trapped-ion quantum computers require light-matter interfaces for the efficient conversion of the quantum state of stationary trapped-ion qubits onto flying photonic qubits. A common path to reaching this goal is to interface the trapped ions with a single mode of an optical cavity \cite{Guthohrlein2001,Mundt2002,Russo2009,Leibrandt2009,Herskind2009,Sterk2012,Stute2012,Bylinskii2015,Steiner2013,Steiner2014,Ballance2017,Takahashi2017,Walker2018}. By employing the principles of cavity quantum electrodynamics, the quantum state of the ion can be mapped onto the photon field.
However, the efficiency of the conversion process is hindered by losses. ``Good'' losses are the residual transmission of the cavity mirrors since the photon eventually escapes the cavity and can be used for transmitting information over long distances. ``Bad'' losses are losses in which the photon is absorbed or scattered out of the cavity mode. In principle, the latter can be minimized by employing high-quality optical coatings, the technology of which has steadily improved over the past decades.

Strong optical transitions of trapped ions usually are in the blue and ultraviolet spectral range. Over the past years, high-finesse cavities for such wavelengths have been developed both using conventional mirrors \cite{Guthohrlein2001,Leibrandt2009,Sterk2012,Bylinskii2015} as well as fiber Fabry-Perot resonators \cite{Ballance2017}. However, in several experiments it was found that intrinsic losses of the cavities increased over time under ultrahigh-vacuum conditions. A possible explanation for this effect was presented in \cite{Gangloff2015}, namely oxygen diffusion out of the top high-index material of the coating. The effect was reversible by exposing the mirrors to air and could, to a good degree, be prevented by placing a few-nanometer thick protective layer of SiO$_2$ on the front surface of the mirror. The reversibility of the degradation by exposition to air was also reported in \cite{Ballance2017}, however, that experiment observed significant degradation despite a protective SiO$_2$ coating.

Here, we follow a different route to address and resolve the problem of UV mirror degradation under ultrahigh vacuum conditions. The work of numerous authors \cite{Hovis1995,Riede2011,Schroeder2013,Wagner2014} suggests that hydrocarbons or other outgassing organic compounds deposit under ultrahigh vacuum conditions on the surface of optical coatings in presence of ultraviolet radiation. The shape of deposition was found to  reflect the laser mode suggesting a laser-induced cracking and deposition. Since ultrahigh vacuum chambers with ion traps and optical cavities often contain  hydrocarbons, for example, as insulating materials for wires and spacers or glue for fixing the mirrors, it is conceivable that UV-enhanced deposition of hydrocarbons contributes also to the widely observed degradation of  UV cavity mirrors. Recovery at air exposure would be a result of cleaning hydrocarbons off the mirror surface by full or partial reaction with oxygen contained in air.

\section{Experimental setup}
In order to test the hypothesis, we have designed and built a novel apparatus containing a linear Paul trap and a high-finesse optical cavity under ultraclean conditions and have completely avoided any organic materials in the construction of the apparatus.

\begin{figure}
	\includegraphics[width=\columnwidth]{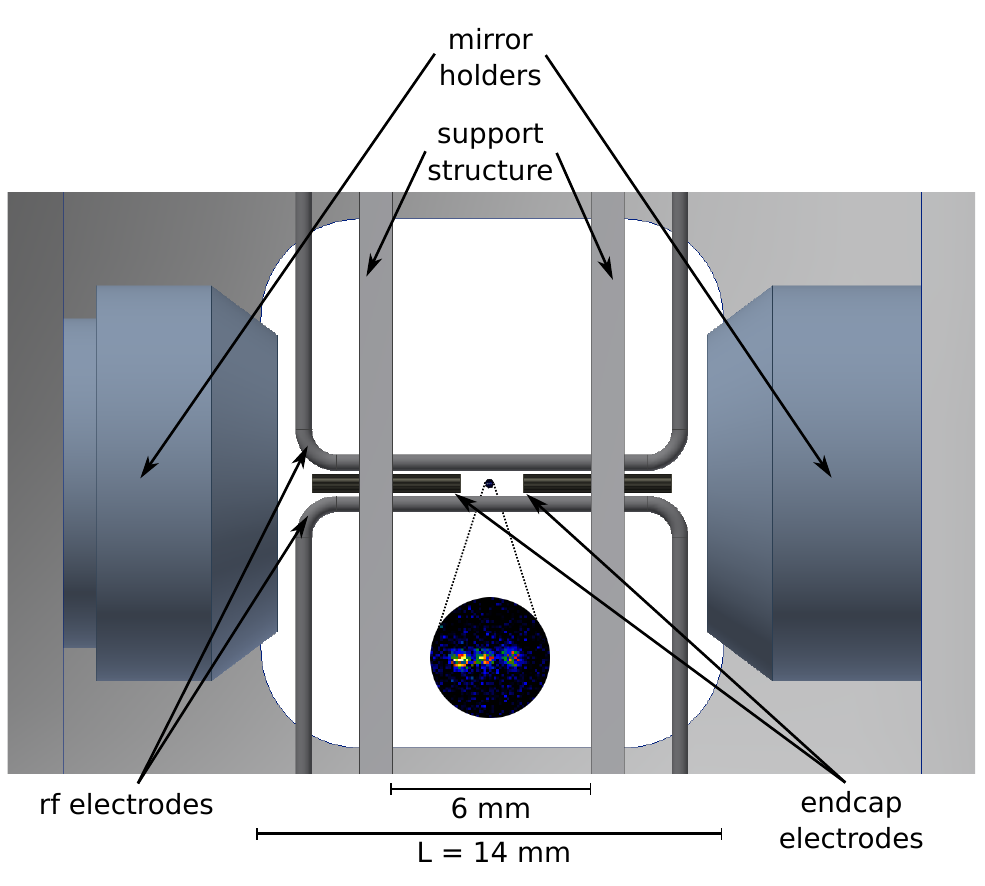}
	\caption{Schematic drawing of the ion trap placed between the cavity mirrors. A radio-frequency field is applied to four rf-electrodes  and axial confinement is provided by hollow endcap electrodes supplied with dc voltages. All electrodes are supported by an Al$_2$O$_3$ structure for electrical isolation. The magnified inset shows a picture of three trapped ions fluorescing at 370\,nm.}
	\label{fig: setup_drawing}
\end{figure}

A schematic overview of the setup is shown in Figure \ref{fig: setup_drawing}. The ion trap is a linear Paul trap with radiofrequency electrodes made from tungsten and an ion-electrode separation of 0.6\,mm. Longitudinal confinement is provided by dc-endcap electrodes made from platinum. The endcap electrodes have a separation of 2.5\,mm and they are hollow with a bore diameter of 0.5\,mm in order to provide space for the field of the optical cavity along the symmetry axis of the Paul trap. The  trapping frequencies for a single Yb$^+$ ion in this trap are $\omega_\perp=2\pi\times 0.49$\,MHz at a driving voltage of 500\,V$_{\text{pp}}$ and $\omega_z=2\pi \times 0.23$\,MHz at an endcap voltage of 10\,V.

The optical cavity comprises of two mirrors with radii of curvature of 5\,cm and a separation of $L=1.4$\,cm. With the speed of light $c$, the corresponding free spectral range is $\nu_{FSR}=c/(2L)=10.7$\,GHz. The optical coatings of the cavity mirrors feature a top layer made of SiO$_2$  having a thickness of $\lambda/(2n)$, where $\lambda=370\,$nm is the employed resonance wavelength of the resonator, and $n\simeq 1.4$ is the refractive index of silicon dioxide. Hence, the top layer of the coating is optically inactive and provides a $>100$\,nm thick protective layer for the mirror.  Nominally, the  coating features a transmission of ${\cal T}=370$\,ppm at 370\,nm and losses of ${\cal L}\lesssim 100$\,ppm.

\begin{figure}[htbp]
	\includegraphics[width=\columnwidth]{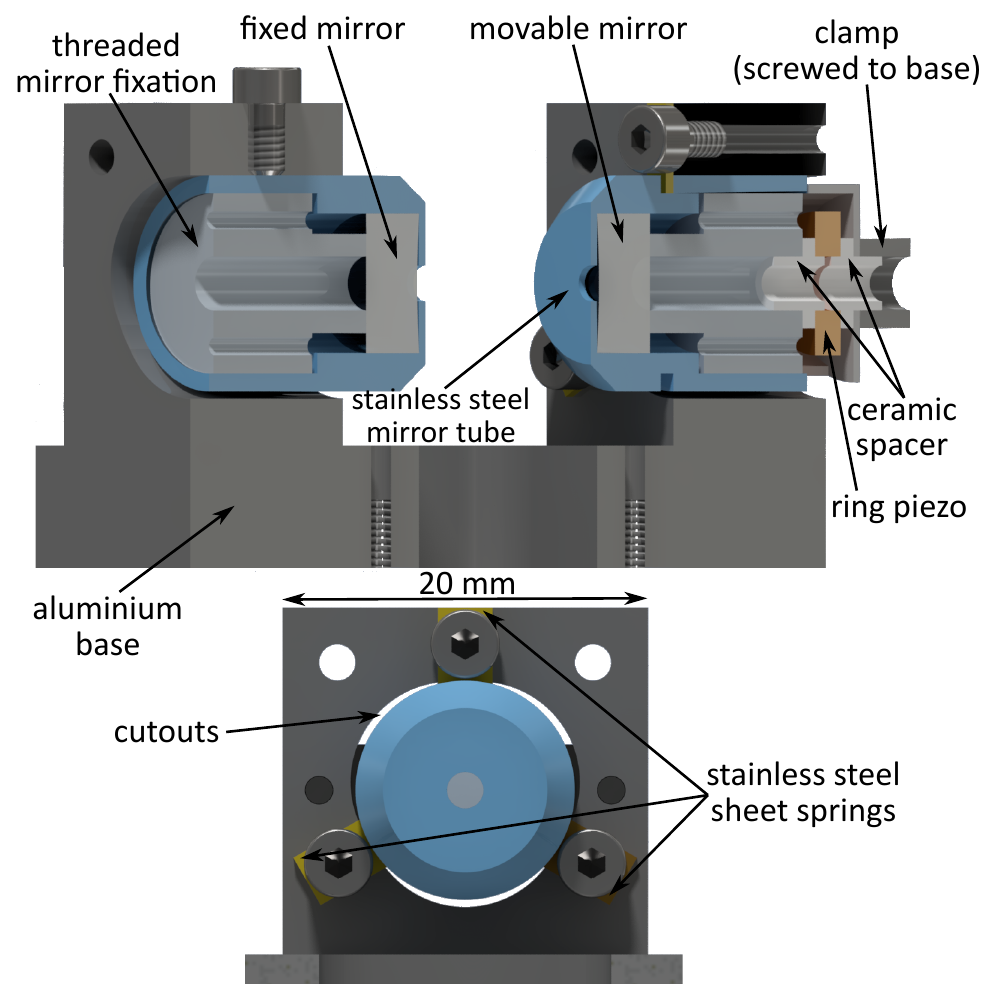}
	\caption{Schematic drawing of the hydrocarbon-free cavity assembly as described in the text.}
	\label{fig: cavity_drawing}
\end{figure}

The optical cavity requires active control of its length in order to tune the resonance wavelength to the resonance frequency of the trapped ions. To this end, we utilize  a piezo-electric transducer to change the position of one of the mirrors. In our setup, we realize this in the following way: The mirror is fixed into a stainless steel tube by a threaded retaining ring (see Figure \ref{fig: cavity_drawing}) and the steel tube serves two purposes: (1) it partially covers the mirror surface in order to prevent charge buildup on the dielectric surface and (2) it acts as a glider for the mirror inside the stainless steel support structure. The contact surface between the glider and the support is minimized by machining a suitably shaped bore into the support, and the glider is fixed to the support structure by three sheet springs mounted at relative angles of 120$^\circ$. The piezo-electric transducer is mechanically clamped between the glider and the support structure and clamping pressure (and hence force pre-loading the piezo transducer) can be  adjusted by varying the thickness of the sheet springs. It should be noted that the constant load on the piezo reduces the available stroke and hence clamping force and piezo is well matched to provide a travel range of the mirror over at least one free spectral range of the cavity.

Supplying both the ion trap and the piezo-electric transducers of the optical cavity with electrical signals requires a significant amount of wiring inside the ultrahigh vacuum chamber. We exclusively use bare copper wires without any insulating coating. Wherever necessary, we added ceramic spacers and tubes in order to ensure physical separation of conductors. Furthermore, unlike in our previous setups \cite{Steiner2013,Ballance2017}, we discarded the passive vibration isolation (by rubber compounds) of the optical cavity inside the vacuum chamber but instead installed vibration damping of the whole ultrahigh-vacuum chamber outside the vacuum. Finally, the ultrahigh vacuum system has been baked at 110$^\circ$C for one week to facilitate outgassing of the vacuum chamber and the installed components. During the bakeout procedure the cavity mirrors had been removed from the setup in order to not contaminate them. The mirrors have been added after the bakeout and subsequently the vacuum system was pumped down to a base pressure of  $\sim 10^{-10}$\,mbar without further baking.

\section{Measurements}

\begin{figure}
	\centering
	\includegraphics[width=\columnwidth]{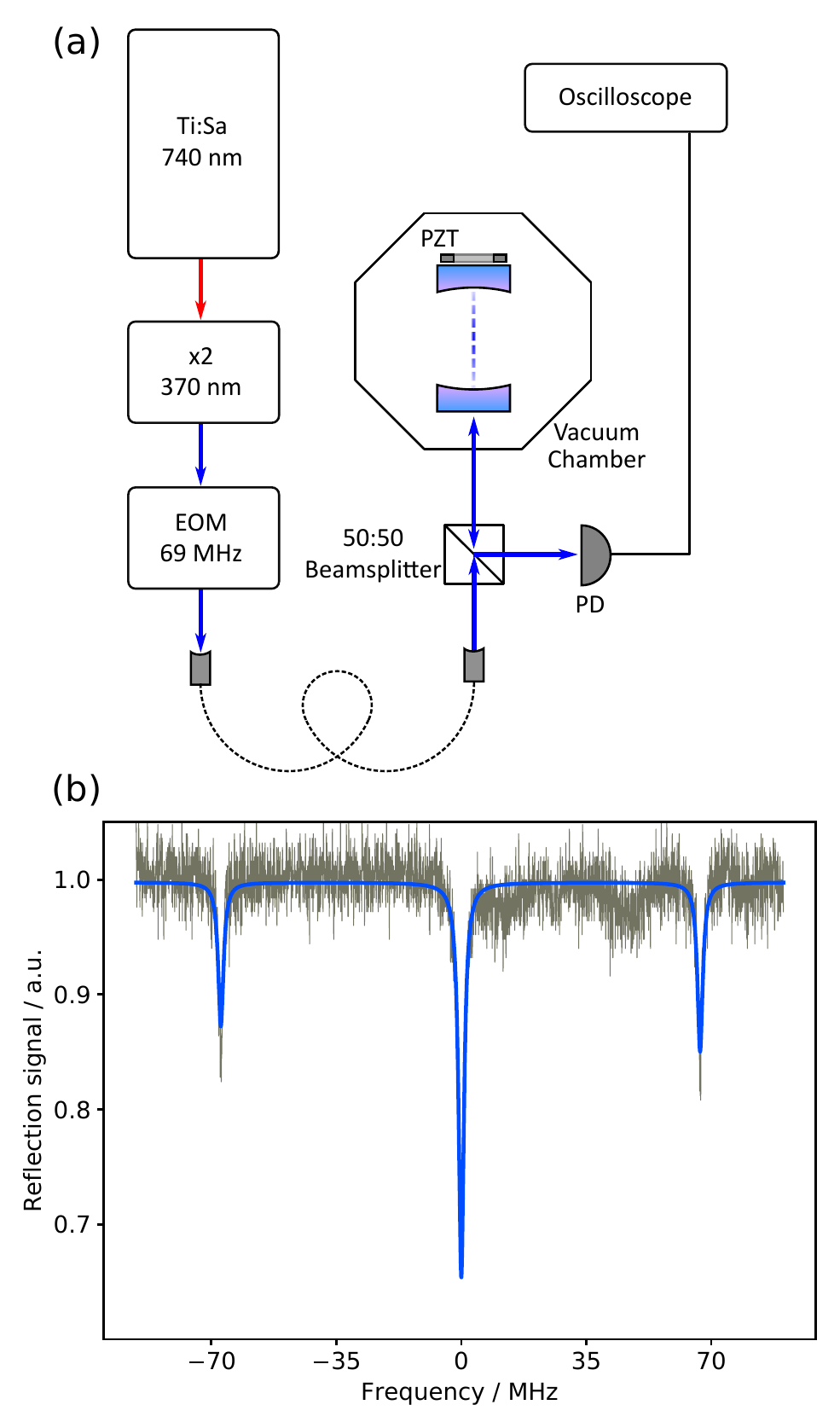}
	\caption{(a) Experimental setup used to determine the mirror losses. The cavity length $L$ is scanned using a piezoelectric transducer (PZT) while the reflection signal is recorded by a photodiode (PD) monitored on an oscilloscope. The resulting spectrum (b) features dips where the cavity is resonant with the incident light. Its linewidth is determined by a lorentzian fit and used to calculate the mirror losses.}
	\label{fig: setup_measurement}
\end{figure}

We have measured the finesse of the cavity (see Figure 3) using a frequency-doubled titanium:sapphire laser with a nominal linewidth of 100\,kHz, much below the cavity linewidth of $\sim 2$\,MHz. Spectroscopy on the Fabry-Perot cavity is performed in reflection. Using an electro-optic modulator, we modulate side bands onto the laser light at $\pm 69$\,MHz, which we use as frequency markers, and we obtain the cavity linewidth $2\kappa$ from a Lorentzian fit to the reflected cavity signal. The finesse is given by $F=\nu_{FSR}/(2\kappa)$.  We determine the mirror losses $\cal{L}$ from the cavity finesse by ${\cal L}=\pi/F-{\cal T}$, assuming equal losses per mirror. For all measurements, the laser power incident on the cavity has been on the order of 80\,$\mu$W. The resulting intracavity intensities are orders of magnitude below the coating damage threshold and thus should not affect the mirror performance even under continuous illumination.

\begin{figure}

	\includegraphics[width=\columnwidth]{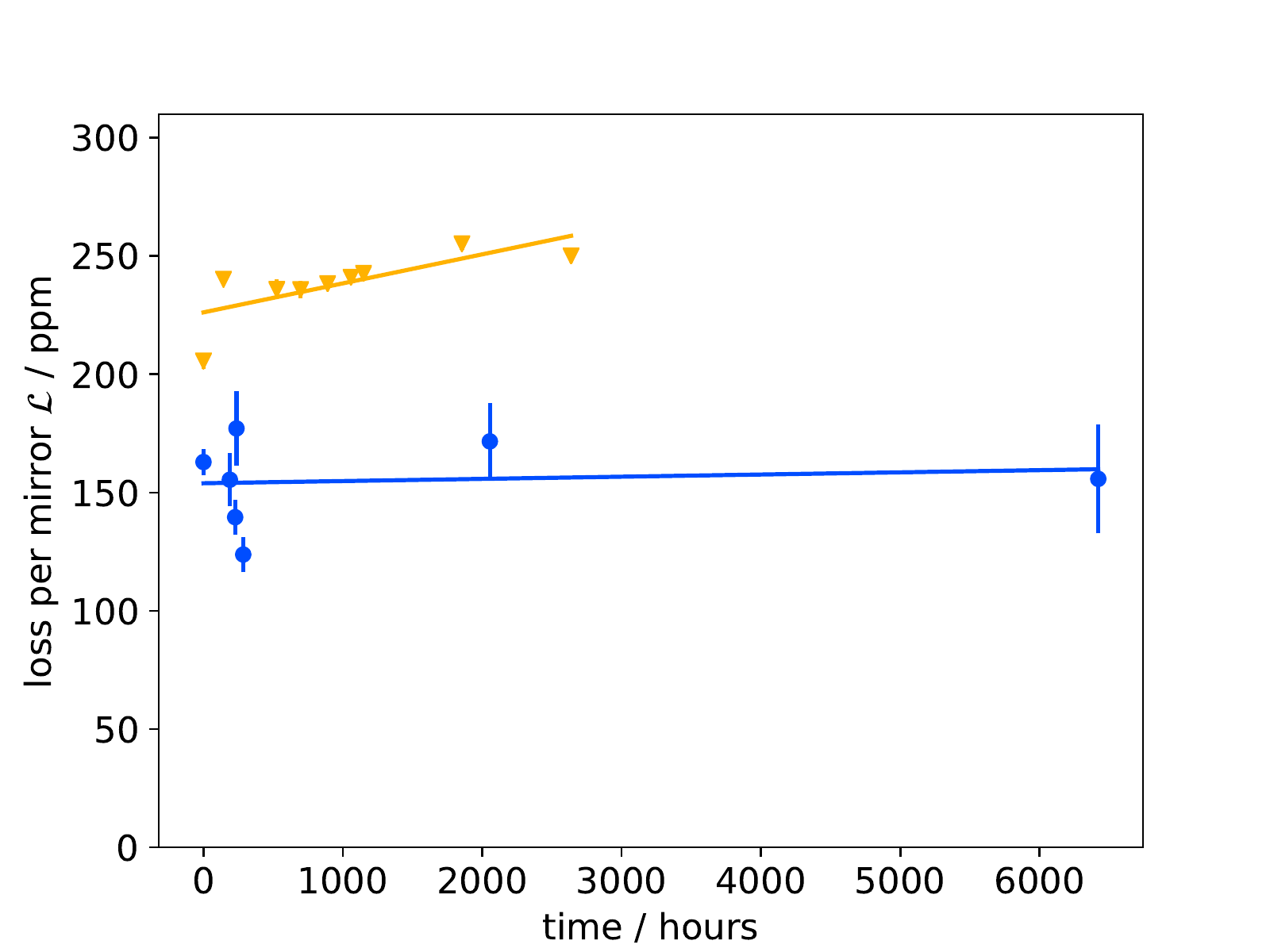}
	\caption{Measurement of cavity losses as a function of time. Our data (blue circles) display no measurable degradation of the mirror losses over a period of 9 months. Each data point is an average of several measurements with statistical errors.  For comparison, we  show the room-temperature data for the same wavelength from Fig. 2a of reference \cite{Gangloff2015} (yellow triangles) together with a linear fit.}
	\label{fig: losses_ybba_vuletic}
\end{figure}

Over a period of 9 months ($\sim 6500$\,hours), we have continually measured the cavity losses while maintaining a vacuum chamber pressure well below $10^{-9}$\,mbar at room temperature.  As shown in Figure \ref{fig: losses_ybba_vuletic}, we did not observe a significant increase of losses on the considered timescale. A linear model yields a variation of $(0.9\pm3.5)\cdot10^{-3}$\,ppm/h, compatible with zero. For comparison,  the smallest degradation rate reported in \cite{Gangloff2015} for the same wavelength was $(12.3\pm4.3)\cdot10^{-3}$\,ppm/h, however, with a Ta$_2$O$_5$ top layer of the mirror coating. The other data point from the literature we can compare to is the rate measured in \cite{Ballance2017} of $(230\pm20)\cdot10^{-3}$\,ppm/h, which is considerably larger. In that work, a 120\,nm thick SiO$_2$ top layer is present on the mirrors, however, some amounts of hydrocarbons (especially glue) have been used in the assembly of the vacuum system and we suspect that they cause the degradation of the cavity. 

In conclusion, we have demonstrated an optical cavity working under ultrahigh-vacuum conditions in the ultraviolet spectral range that operates without degradation for 9 months. By comparison with previous data for the same spectral range \cite{Gangloff2015,Ballance2017} our experimental approach to remove hydrocarbons from the vacuum system seems to yield a significant improve in the stability of then cavity finesse over time.

We thank T. Gross for valuable discussions, and D. Gangloff and V. Vuletic for sharing their data. This work was supported by DFG (SFB/TR 185, A2), Cluster of Excellence Matter and Light for Quantum Computing (ML4Q) EXC 2004/1 – 390534769, BMBF (FaResQ), and the Alexander-von-Humboldt Stiftung.


\begin{thebibliography}{18}%
\makeatletter
\providecommand \@ifxundefined [1]{%
 \@ifx{#1\undefined}
}%
\providecommand \@ifnum [1]{%
 \ifnum #1\expandafter \@firstoftwo
 \else \expandafter \@secondoftwo
 \fi
}%
\providecommand \@ifx [1]{%
 \ifx #1\expandafter \@firstoftwo
 \else \expandafter \@secondoftwo
 \fi
}%
\providecommand \natexlab [1]{#1}%
\providecommand \enquote  [1]{``#1''}%
\providecommand \bibnamefont  [1]{#1}%
\providecommand \bibfnamefont [1]{#1}%
\providecommand \citenamefont [1]{#1}%
\providecommand \href@noop [0]{\@secondoftwo}%
\providecommand \href [0]{\begingroup \@sanitize@url \@href}%
\providecommand \@href[1]{\@@startlink{#1}\@@href}%
\providecommand \@@href[1]{\endgroup#1\@@endlink}%
\providecommand \@sanitize@url [0]{\catcode `\\12\catcode `\$12\catcode
  `\&12\catcode `\#12\catcode `\^12\catcode `\_12\catcode `\%12\relax}%
\providecommand \@@startlink[1]{}%
\providecommand \@@endlink[0]{}%
\providecommand \url  [0]{\begingroup\@sanitize@url \@url }%
\providecommand \@url [1]{\endgroup\@href {#1}{\urlprefix }}%
\providecommand \urlprefix  [0]{URL }%
\providecommand \Eprint [0]{\href }%
\providecommand \doibase [0]{http://dx.doi.org/}%
\providecommand \selectlanguage [0]{\@gobble}%
\providecommand \bibinfo  [0]{\@secondoftwo}%
\providecommand \bibfield  [0]{\@secondoftwo}%
\providecommand \translation [1]{[#1]}%
\providecommand \BibitemOpen [0]{}%
\providecommand \bibitemStop [0]{}%
\providecommand \bibitemNoStop [0]{.\EOS\space}%
\providecommand \EOS [0]{\spacefactor3000\relax}%
\providecommand \BibitemShut  [1]{\csname bibitem#1\endcsname}%
\let\auto@bib@innerbib\@empty
\bibitem [{\citenamefont {Guth{\"o}hrlein}\ \emph {et~al.}(2001)\citenamefont
  {Guth{\"o}hrlein}, \citenamefont {Keller}, \citenamefont {Hayasaka},
  \citenamefont {Lange},\ and\ \citenamefont {Walther}}]{Guthohrlein2001}%
  \BibitemOpen
  \bibfield  {author} {\bibinfo {author} {\bibfnamefont {G.~R.}\ \bibnamefont
  {Guth{\"o}hrlein}}, \bibinfo {author} {\bibfnamefont {M.}~\bibnamefont
  {Keller}}, \bibinfo {author} {\bibfnamefont {K.}~\bibnamefont {Hayasaka}},
  \bibinfo {author} {\bibfnamefont {W.}~\bibnamefont {Lange}}, \ and\ \bibinfo
  {author} {\bibfnamefont {H.}~\bibnamefont {Walther}},\ }\href
  {http://dx.doi.org/10.1038/35102129} {\bibfield  {journal} {\bibinfo
  {journal} {Nature}\ }\textbf {\bibinfo {volume} {414}},\ \bibinfo {pages}
  {49} (\bibinfo {year} {2001})}\BibitemShut {NoStop}%
\bibitem [{\citenamefont {Mundt}\ \emph {et~al.}(2002)\citenamefont {Mundt},
  \citenamefont {Kreuter}, \citenamefont {Becher}, \citenamefont {Leibfried},
  \citenamefont {Eschner}, \citenamefont {Schmidt-Kaler},\ and\ \citenamefont
  {Blatt}}]{Mundt2002}%
  \BibitemOpen
  \bibfield  {author} {\bibinfo {author} {\bibfnamefont {A.~B.}\ \bibnamefont
  {Mundt}}, \bibinfo {author} {\bibfnamefont {A.}~\bibnamefont {Kreuter}},
  \bibinfo {author} {\bibfnamefont {C.}~\bibnamefont {Becher}}, \bibinfo
  {author} {\bibfnamefont {D.}~\bibnamefont {Leibfried}}, \bibinfo {author}
  {\bibfnamefont {J.}~\bibnamefont {Eschner}}, \bibinfo {author} {\bibfnamefont
  {F.}~\bibnamefont {Schmidt-Kaler}}, \ and\ \bibinfo {author} {\bibfnamefont
  {R.}~\bibnamefont {Blatt}},\ }\href {\doibase 10.1103/PhysRevLett.89.103001}
  {\bibfield  {journal} {\bibinfo  {journal} {Phys. Rev. Lett.}\ }\textbf
  {\bibinfo {volume} {89}},\ \bibinfo {pages} {103001} (\bibinfo {year}
  {2002})}\BibitemShut {NoStop}%
\bibitem [{\citenamefont {Russo}\ \emph {et~al.}(2009)\citenamefont {Russo},
  \citenamefont {Barros}, \citenamefont {Stute}, \citenamefont {Dubin},
  \citenamefont {Phillips}, \citenamefont {Monz}, \citenamefont {Northup},
  \citenamefont {Becher}, \citenamefont {Salzburger}, \citenamefont {Ritsch},
  \citenamefont {Schmidt},\ and\ \citenamefont {Blatt}}]{Russo2009}%
  \BibitemOpen
  \bibfield  {author} {\bibinfo {author} {\bibfnamefont {C.}~\bibnamefont
  {Russo}}, \bibinfo {author} {\bibfnamefont {H.}~\bibnamefont {Barros}},
  \bibinfo {author} {\bibfnamefont {A.}~\bibnamefont {Stute}}, \bibinfo
  {author} {\bibfnamefont {F.}~\bibnamefont {Dubin}}, \bibinfo {author}
  {\bibfnamefont {E.}~\bibnamefont {Phillips}}, \bibinfo {author}
  {\bibfnamefont {T.}~\bibnamefont {Monz}}, \bibinfo {author} {\bibfnamefont
  {T.}~\bibnamefont {Northup}}, \bibinfo {author} {\bibfnamefont
  {C.}~\bibnamefont {Becher}}, \bibinfo {author} {\bibfnamefont
  {T.}~\bibnamefont {Salzburger}}, \bibinfo {author} {\bibfnamefont
  {H.}~\bibnamefont {Ritsch}}, \bibinfo {author} {\bibfnamefont
  {P.}~\bibnamefont {Schmidt}}, \ and\ \bibinfo {author} {\bibfnamefont
  {R.}~\bibnamefont {Blatt}},\ }\href {\doibase 10.1007/s00340-009-3430-2}
  {\bibfield  {journal} {\bibinfo  {journal} {Applied Physics B}\ }\textbf
  {\bibinfo {volume} {95}},\ \bibinfo {pages} {205} (\bibinfo {year}
  {2009})}\BibitemShut {NoStop}%
\bibitem [{\citenamefont {Leibrandt}\ \emph {et~al.}(2009)\citenamefont
  {Leibrandt}, \citenamefont {Labaziewicz}, \citenamefont
  {Vuleti\ifmmode~\acute{c}\else \'{c}\fi{}},\ and\ \citenamefont
  {Chuang}}]{Leibrandt2009}%
  \BibitemOpen
  \bibfield  {author} {\bibinfo {author} {\bibfnamefont {D.~R.}\ \bibnamefont
  {Leibrandt}}, \bibinfo {author} {\bibfnamefont {J.}~\bibnamefont
  {Labaziewicz}}, \bibinfo {author} {\bibfnamefont {V.}~\bibnamefont
  {Vuleti\ifmmode~\acute{c}\else \'{c}\fi{}}}, \ and\ \bibinfo {author}
  {\bibfnamefont {I.~L.}\ \bibnamefont {Chuang}},\ }\href {\doibase
  10.1103/PhysRevLett.103.103001} {\bibfield  {journal} {\bibinfo  {journal}
  {Phys. Rev. Lett.}\ }\textbf {\bibinfo {volume} {103}},\ \bibinfo {pages}
  {103001} (\bibinfo {year} {2009})}\BibitemShut {NoStop}%
\bibitem [{\citenamefont {Herskind}\ \emph {et~al.}(2009)\citenamefont
  {Herskind}, \citenamefont {Dantan}, \citenamefont {Marler}, \citenamefont
  {Albert},\ and\ \citenamefont {Drewsen}}]{Herskind2009}%
  \BibitemOpen
  \bibfield  {author} {\bibinfo {author} {\bibfnamefont {P.~F.}\ \bibnamefont
  {Herskind}}, \bibinfo {author} {\bibfnamefont {A.}~\bibnamefont {Dantan}},
  \bibinfo {author} {\bibfnamefont {J.~P.}\ \bibnamefont {Marler}}, \bibinfo
  {author} {\bibfnamefont {M.}~\bibnamefont {Albert}}, \ and\ \bibinfo {author}
  {\bibfnamefont {M.}~\bibnamefont {Drewsen}},\ }\href
  {http://dx.doi.org/10.1038/nphys1302} {\bibfield  {journal} {\bibinfo
  {journal} {Nature Physics}\ }\textbf {\bibinfo {volume} {5}},\ \bibinfo
  {pages} {494 EP } (\bibinfo {year} {2009})}\BibitemShut {NoStop}%
\bibitem [{\citenamefont {Sterk}\ \emph {et~al.}(2012)\citenamefont {Sterk},
  \citenamefont {Luo}, \citenamefont {Manning}, \citenamefont {Maunz},\ and\
  \citenamefont {Monroe}}]{Sterk2012}%
  \BibitemOpen
  \bibfield  {author} {\bibinfo {author} {\bibfnamefont {J.~D.}\ \bibnamefont
  {Sterk}}, \bibinfo {author} {\bibfnamefont {L.}~\bibnamefont {Luo}}, \bibinfo
  {author} {\bibfnamefont {T.~A.}\ \bibnamefont {Manning}}, \bibinfo {author}
  {\bibfnamefont {P.}~\bibnamefont {Maunz}}, \ and\ \bibinfo {author}
  {\bibfnamefont {C.}~\bibnamefont {Monroe}},\ }\href {\doibase
  10.1103/PhysRevA.85.062308} {\bibfield  {journal} {\bibinfo  {journal} {Phys.
  Rev. A}\ }\textbf {\bibinfo {volume} {85}},\ \bibinfo {pages} {062308}
  (\bibinfo {year} {2012})}\BibitemShut {NoStop}%
\bibitem [{\citenamefont {Stute}\ \emph {et~al.}(2012)\citenamefont {Stute},
  \citenamefont {Casabone}, \citenamefont {Schindler}, \citenamefont {Monz},
  \citenamefont {Schmidt}, \citenamefont {Brandstatter}, \citenamefont
  {Northup},\ and\ \citenamefont {Blatt}}]{Stute2012}%
  \BibitemOpen
  \bibfield  {author} {\bibinfo {author} {\bibfnamefont {A.}~\bibnamefont
  {Stute}}, \bibinfo {author} {\bibfnamefont {B.}~\bibnamefont {Casabone}},
  \bibinfo {author} {\bibfnamefont {P.}~\bibnamefont {Schindler}}, \bibinfo
  {author} {\bibfnamefont {T.}~\bibnamefont {Monz}}, \bibinfo {author}
  {\bibfnamefont {P.~O.}\ \bibnamefont {Schmidt}}, \bibinfo {author}
  {\bibfnamefont {B.}~\bibnamefont {Brandstatter}}, \bibinfo {author}
  {\bibfnamefont {T.~E.}\ \bibnamefont {Northup}}, \ and\ \bibinfo {author}
  {\bibfnamefont {R.}~\bibnamefont {Blatt}},\ }\href
  {http://dx.doi.org/10.1038/nature11120} {\bibfield  {journal} {\bibinfo
  {journal} {Nature}\ }\textbf {\bibinfo {volume} {485}},\ \bibinfo {pages}
  {482} (\bibinfo {year} {2012})}\BibitemShut {NoStop}%
\bibitem [{\citenamefont {Bylinskii}\ \emph {et~al.}(2015)\citenamefont
  {Bylinskii}, \citenamefont {Gangloff},\ and\ \citenamefont
  {Vuleti{\'c}}}]{Bylinskii2015}%
  \BibitemOpen
  \bibfield  {author} {\bibinfo {author} {\bibfnamefont {A.}~\bibnamefont
  {Bylinskii}}, \bibinfo {author} {\bibfnamefont {D.}~\bibnamefont {Gangloff}},
  \ and\ \bibinfo {author} {\bibfnamefont {V.}~\bibnamefont {Vuleti{\'c}}},\
  }\href {\doibase 10.1126/science.1261422} {\bibfield  {journal} {\bibinfo
  {journal} {Science}\ }\textbf {\bibinfo {volume} {348}},\ \bibinfo {pages}
  {1115} (\bibinfo {year} {2015})}\  \BibitemShut
  {NoStop}%
\bibitem [{\citenamefont {Steiner}\ \emph {et~al.}(2013)\citenamefont
  {Steiner}, \citenamefont {Meyer}, \citenamefont {Deutsch}, \citenamefont
  {Reichel},\ and\ \citenamefont {K\"ohl}}]{Steiner2013}%
  \BibitemOpen
  \bibfield  {author} {\bibinfo {author} {\bibfnamefont {M.}~\bibnamefont
  {Steiner}}, \bibinfo {author} {\bibfnamefont {H.~M.}\ \bibnamefont {Meyer}},
  \bibinfo {author} {\bibfnamefont {C.}~\bibnamefont {Deutsch}}, \bibinfo
  {author} {\bibfnamefont {J.}~\bibnamefont {Reichel}}, \ and\ \bibinfo
  {author} {\bibfnamefont {M.}~\bibnamefont {K\"ohl}},\ }\href {\doibase
  10.1103/PhysRevLett.110.043003} {\bibfield  {journal} {\bibinfo  {journal}
  {Phys. Rev. Lett.}\ }\textbf {\bibinfo {volume} {110}},\ \bibinfo {pages}
  {043003} (\bibinfo {year} {2013})}\BibitemShut {NoStop}%
\bibitem [{\citenamefont {Steiner}\ \emph {et~al.}(2014)\citenamefont
  {Steiner}, \citenamefont {Meyer}, \citenamefont {Reichel},\ and\
  \citenamefont {K\"ohl}}]{Steiner2014}%
  \BibitemOpen
  \bibfield  {author} {\bibinfo {author} {\bibfnamefont {M.}~\bibnamefont
  {Steiner}}, \bibinfo {author} {\bibfnamefont {H.~M.}\ \bibnamefont {Meyer}},
  \bibinfo {author} {\bibfnamefont {J.}~\bibnamefont {Reichel}}, \ and\
  \bibinfo {author} {\bibfnamefont {M.}~\bibnamefont {K\"ohl}},\ }\href
  {\doibase 10.1103/PhysRevLett.113.263003} {\bibfield  {journal} {\bibinfo
  {journal} {Phys. Rev. Lett.}\ }\textbf {\bibinfo {volume} {113}},\ \bibinfo
  {pages} {263003} (\bibinfo {year} {2014})}\BibitemShut {NoStop}%
\bibitem [{\citenamefont {Ballance}\ \emph {et~al.}(2017)\citenamefont
  {Ballance}, \citenamefont {Meyer}, \citenamefont {Kobel}, \citenamefont
  {Ott}, \citenamefont {Reichel},\ and\ \citenamefont {K\"ohl}}]{Ballance2017}%
  \BibitemOpen
  \bibfield  {author} {\bibinfo {author} {\bibfnamefont {T.~G.}\ \bibnamefont
  {Ballance}}, \bibinfo {author} {\bibfnamefont {H.~M.}\ \bibnamefont {Meyer}},
  \bibinfo {author} {\bibfnamefont {P.}~\bibnamefont {Kobel}}, \bibinfo
  {author} {\bibfnamefont {K.}~\bibnamefont {Ott}}, \bibinfo {author}
  {\bibfnamefont {J.}~\bibnamefont {Reichel}}, \ and\ \bibinfo {author}
  {\bibfnamefont {M.}~\bibnamefont {K\"ohl}},\ }\href {\doibase
  10.1103/PhysRevA.95.033812} {\bibfield  {journal} {\bibinfo  {journal} {Phys.
  Rev. A}\ }\textbf {\bibinfo {volume} {95}},\ \bibinfo {pages} {033812}
  (\bibinfo {year} {2017})}\BibitemShut {NoStop}%
\bibitem [{\citenamefont {Takahashi}\ \emph {et~al.}(2017)\citenamefont
  {Takahashi}, \citenamefont {Kassa}, \citenamefont {Christoforou},\ and\
  \citenamefont {Keller}}]{Takahashi2017}%
  \BibitemOpen
  \bibfield  {author} {\bibinfo {author} {\bibfnamefont {H.}~\bibnamefont
  {Takahashi}}, \bibinfo {author} {\bibfnamefont {E.}~\bibnamefont {Kassa}},
  \bibinfo {author} {\bibfnamefont {C.}~\bibnamefont {Christoforou}}, \ and\
  \bibinfo {author} {\bibfnamefont {M.}~\bibnamefont {Keller}},\ }\href
  {\doibase 10.1103/PhysRevA.96.023824} {\bibfield  {journal} {\bibinfo
  {journal} {Phys. Rev. A}\ }\textbf {\bibinfo {volume} {96}},\ \bibinfo
  {pages} {023824} (\bibinfo {year} {2017})}\BibitemShut {NoStop}%
\bibitem [{\citenamefont {Walker}\ \emph {et~al.}(2018)\citenamefont {Walker},
  \citenamefont {Miyanishi}, \citenamefont {Ikuta}, \citenamefont {Takahashi},
  \citenamefont {Vartabi~Kashanian}, \citenamefont {Tsujimoto}, \citenamefont
  {Hayasaka}, \citenamefont {Yamamoto}, \citenamefont {Imoto},\ and\
  \citenamefont {Keller}}]{Walker2018}%
  \BibitemOpen
  \bibfield  {author} {\bibinfo {author} {\bibfnamefont {T.}~\bibnamefont
  {Walker}}, \bibinfo {author} {\bibfnamefont {K.}~\bibnamefont {Miyanishi}},
  \bibinfo {author} {\bibfnamefont {R.}~\bibnamefont {Ikuta}}, \bibinfo
  {author} {\bibfnamefont {H.}~\bibnamefont {Takahashi}}, \bibinfo {author}
  {\bibfnamefont {S.}~\bibnamefont {Vartabi~Kashanian}}, \bibinfo {author}
  {\bibfnamefont {Y.}~\bibnamefont {Tsujimoto}}, \bibinfo {author}
  {\bibfnamefont {K.}~\bibnamefont {Hayasaka}}, \bibinfo {author}
  {\bibfnamefont {T.}~\bibnamefont {Yamamoto}}, \bibinfo {author}
  {\bibfnamefont {N.}~\bibnamefont {Imoto}}, \ and\ \bibinfo {author}
  {\bibfnamefont {M.}~\bibnamefont {Keller}},\ }\href {\doibase
  10.1103/PhysRevLett.120.203601} {\bibfield  {journal} {\bibinfo  {journal}
  {Phys. Rev. Lett.}\ }\textbf {\bibinfo {volume} {120}},\ \bibinfo {pages}
  {203601} (\bibinfo {year} {2018})}\BibitemShut {NoStop}%
\bibitem [{\citenamefont {Gangloff}\ \emph {et~al.}(2015)\citenamefont
  {Gangloff}, \citenamefont {Shi}, \citenamefont {Wu}, \citenamefont
  {Bylinskii}, \citenamefont {Braverman}, \citenamefont {Gutierrez},
  \citenamefont {Nichols}, \citenamefont {Li}, \citenamefont {Aichholz},
  \citenamefont {Cetina}, \citenamefont {Karpa}, \citenamefont
  {Jelenkovi\'{c}}, \citenamefont {Chuang},\ and\ \citenamefont
  {Vuleti\'{c}}}]{Gangloff2015}%
  \BibitemOpen
  \bibfield  {author} {\bibinfo {author} {\bibfnamefont {D.}~\bibnamefont
  {Gangloff}}, \bibinfo {author} {\bibfnamefont {M.}~\bibnamefont {Shi}},
  \bibinfo {author} {\bibfnamefont {T.}~\bibnamefont {Wu}}, \bibinfo {author}
  {\bibfnamefont {A.}~\bibnamefont {Bylinskii}}, \bibinfo {author}
  {\bibfnamefont {B.}~\bibnamefont {Braverman}}, \bibinfo {author}
  {\bibfnamefont {M.}~\bibnamefont {Gutierrez}}, \bibinfo {author}
  {\bibfnamefont {R.}~\bibnamefont {Nichols}}, \bibinfo {author} {\bibfnamefont
  {J.}~\bibnamefont {Li}}, \bibinfo {author} {\bibfnamefont {K.}~\bibnamefont
  {Aichholz}}, \bibinfo {author} {\bibfnamefont {M.}~\bibnamefont {Cetina}},
  \bibinfo {author} {\bibfnamefont {L.}~\bibnamefont {Karpa}}, \bibinfo
  {author} {\bibfnamefont {B.}~\bibnamefont {Jelenkovi\'{c}}}, \bibinfo
  {author} {\bibfnamefont {I.}~\bibnamefont {Chuang}}, \ and\ \bibinfo {author}
  {\bibfnamefont {V.}~\bibnamefont {Vuleti\'{c}}},\ }\href {\doibase
  10.1364/OE.23.018014} {\bibfield  {journal} {\bibinfo  {journal} {Opt.
  Express}\ }\textbf {\bibinfo {volume} {23}},\ \bibinfo {pages} {18014}
  (\bibinfo {year} {2015})}\BibitemShut {NoStop}%
\bibitem [{\citenamefont {Hovis}\ \emph {et~al.}(1995)\citenamefont {Hovis},
  \citenamefont {Shepherd}, \citenamefont {Radcliffe},\ and\ \citenamefont
  {Maliborski}}]{Hovis1995}%
  \BibitemOpen
  \bibfield  {author} {\bibinfo {author} {\bibfnamefont {F.~E.}\ \bibnamefont
  {Hovis}}, \bibinfo {author} {\bibfnamefont {B.~A.}\ \bibnamefont {Shepherd}},
  \bibinfo {author} {\bibfnamefont {C.~T.}\ \bibnamefont {Radcliffe}}, \ and\
  \bibinfo {author} {\bibfnamefont {H.~A.}\ \bibnamefont {Maliborski}},\ }in\
  \href {\doibase 10.1117/12.213736} {\emph {\bibinfo {booktitle}
  {Laser-Induced Damage in Optical Materials: 1994}}},\ Vol.\ \bibinfo {volume}
  {2428},\ \bibinfo {editor} {edited by\ \bibinfo {editor} {\bibfnamefont
  {H.~E.}\ \bibnamefont {Bennett}}, \bibinfo {editor} {\bibfnamefont {A.~H.}\
  \bibnamefont {Guenther}}, \bibinfo {editor} {\bibfnamefont {M.~R.}\
  \bibnamefont {Kozlowski}}, \bibinfo {editor} {\bibfnamefont {B.~E.}\
  \bibnamefont {Newnam}}, \ and\ \bibinfo {editor} {\bibfnamefont {M.~J.}\
  \bibnamefont {Soileau}}}\ (\bibinfo  {publisher} {{SPIE}},\ \bibinfo {year}
  {1995})\ pp.\ \bibinfo {pages} {2428 -- 2428 -- 12}\BibitemShut {NoStop}%
\bibitem [{\citenamefont {Riede}\ \emph {et~al.}(2012)\citenamefont {Riede},
  \citenamefont {Schr{\"o}der},\ and\ \citenamefont {Wernham}}]{Riede2011}%
  \BibitemOpen
  \bibfield  {author} {\bibinfo {author} {\bibfnamefont {W.}~\bibnamefont
  {Riede}}, \bibinfo {author} {\bibfnamefont {H.}~\bibnamefont {Schr{\"o}der}},
  \ and\ \bibinfo {author} {\bibfnamefont {D.}~\bibnamefont {Wernham}},\ }in\
  \href {https://elib.dlr.de/70941/} {\emph {\bibinfo {booktitle} {Laser Damage
  2011}}},\ \bibinfo {series and number} {Laser Damage},\ \bibinfo {editor}
  {edited by\ \bibinfo {editor} {\bibnamefont {SPIE}}}\ (\bibinfo  {publisher}
  {Society of Photo-Optical Instrumentation Engineers (SPIE).},\ \bibinfo
  {year} {2012})\BibitemShut {NoStop}%
\bibitem [{\citenamefont {Schr{\"o}der}\ \emph {et~al.}(2013)\citenamefont
  {Schr{\"o}der}, \citenamefont {Wagner}, \citenamefont {Kokkinos}, \citenamefont
  {Riede},\ and\ \citenamefont {Tighe}}]{Schroeder2013}%
  \BibitemOpen
  \bibfield  {author} {\bibinfo {author} {\bibfnamefont {H.}~\bibnamefont
  {Schr{\"o}der}}, \bibinfo {author} {\bibfnamefont {P.}~\bibnamefont {Wagner}},
  \bibinfo {author} {\bibfnamefont {D.}~\bibnamefont {Kokkinos}}, \bibinfo
  {author} {\bibfnamefont {W.}~\bibnamefont {Riede}}, \ and\ \bibinfo {author}
  {\bibfnamefont {A.}~\bibnamefont {Tighe}},\ }in\ \href {\doibase
  10.1117/12.2030002} {\emph {\bibinfo {booktitle} {Laser-Induced Damage in
  Optical Materials: 2013}}},\ Vol.\ \bibinfo {volume} {8885},\ \bibinfo
  {editor} {edited by\ \bibinfo {editor} {\bibfnamefont {G.~J.}\ \bibnamefont
  {Exarhos}}, \bibinfo {editor} {\bibfnamefont {V.~E.}\ \bibnamefont
  {Gruzdev}}, \bibinfo {editor} {\bibfnamefont {J.~A.}\ \bibnamefont
  {Menapace}}, \bibinfo {editor} {\bibfnamefont {D.}~\bibnamefont {Ristau}}, \
  and\ \bibinfo {editor} {\bibfnamefont {M.}~\bibnamefont {Soileau}}}\
  (\bibinfo  {publisher} {{SPIE}},\ \bibinfo {year} {2013})\ pp.\ \bibinfo
  {pages} {8885 -- 8885 -- 9}\BibitemShut {NoStop}%
\bibitem [{\citenamefont {Wagner}(2014)}]{Wagner2014}%
  \BibitemOpen
  \bibfield  {author} {\bibinfo {author} {\bibfnamefont {P.}~\bibnamefont
  {Wagner}},\ }\href {https://elib.dlr.de/88540/} {\enquote {\bibinfo {title}
  {Laser-induced contamination on high-reflective optics},}\ } (\bibinfo {year}
  {2014})\BibitemShut {NoStop}%
\end{thebibliography}
\end{document}